\documentclass[a4paper,11pt]{article}

\usepackage{amsmath}
\usepackage{graphicx}
\usepackage{amsfonts}
\usepackage{amssymb}
\usepackage{color}
\usepackage{comment}

\newcommand{\bc}{\begin{center}}
\newcommand{\ec}{\end{center}}
\def\ba#1{\begin{array}{#1}\displaystyle}
\newcommand{\ea}{\end{array}}

\newcommand{\beq}{\begin{equation}}
\newcommand{\eeq}{\end{equation}}
\newcommand{\beqa}{\begin{eqnarray}}
\newcommand{\eeqa}{\end{eqnarray}}

\newcommand{\n}{\nonumber\\}
\newcommand{\bi}{\begin{itemize}}
\newcommand{\ei}{\end{itemize}}
\def\mato{\left(\begin{matrix}} 
\def\matf{\end{matrix}\right)}

\newtheorem{rema}{Remark}[section]

\def\lt#1{\left#1}
\def\rt#1{\right#1}
\def\t#1{\tilde{#1}}

\def\b#1{\bar{#1}}
\def\frc#1#2{\frac{#1}{#2}}

\newcommand{\p}{\partial}

\newcommand{\bra}{\langle}
\newcommand{\ket}{\rangle}

\newcommand{\Or}{{\cal O}}

\newcommand{\ep}{\epsilon}
\newcommand{\varep}{\varepsilon}

\newcommand{\Tr}{{\rm Tr}}

\newcommand{\dd}{{\rm d}}

\newcommand{\uu}{{\tt k}}
\newcommand{\hh}{{\tt h}}
\newcommand{\pp}{{\tt p}}
\newcommand{\jj}{{\tt j}}

\newcommand{\TT}{{\tt T}}
\def\gen#1{{\bra #1\ket}}
\def\genc#1{\langle #1\rangle^{{\rm c}}}

\begin{document}

\begin{center}
{\Large {\bf A Hydrodynamic Approach to\\ \vskip 0.2 truecm Non-Equilibrium Conformal Field Theories}}
\vskip 1.5
 truecm
{\large Denis Bernard${}^{\clubsuit}$ and Benjamin Doyon${}^{\spadesuit}$}

\vspace{0.5cm}
{\small ${}^{\clubsuit}$ Laboratoire de Physique Th\'eorique de l'ENS, CNRS $\&$ Ecole Normale Sup\'erieure de Paris, France.}\\
{\small ${}^{\spadesuit}$ Department of Mathematics, King's College London, London, United Kingdom.}\\
\end{center}

{\centerline \today}

\vspace{1.0cm} 
\noindent{\bf Abstract}\\
We develop a hydrodynamic approach to non-equilibrium conformal field theory. We study non-equilibrium steady states in the context of one-dimensional conformal field theory perturbed by the $T\b T$ irrelevant operator. By direct quantum computation, we show, to first order in the coupling, that a relativistic hydrodynamic emerges, which is a simple modification of one-dimensional conformal fluids. We show that it describes the steady state and its approach, and we provide the main characteristics of the steady state, which lies between two shock waves. The velocities of these shocks are modified by the perturbation and equal the sound velocities of the asymptotic baths. Pushing further this approach, we are led to conjecture that the approach to the steady state is generically controlled by the power law $t^{-1/2}$, and that the widths of the shocks increase with time according to $t^{1/3}$.
\vskip 1.5 truecm

\section{Introduction}

One of the most powerful ideas in studying the dynamics of quantum field theory is that emerging from a hydrodynamic description of local averages \cite{Landau}. Hydrodynamics allows to encode in a simple way non-equilibrium states, including states with constant flows and approaches to steady states, by concentrating only on quantities of physical relevance and without having to deal explicitly with an infinite number of degrees of freedom. In the quantum context, the passage from strongly interacting many-body quantum dynamics to classical hydrodynamics involve subtle effects that are sometimes hard to control \cite{Qhydro}. 
It is therefore important to have simple models and setups where this passage can be studied with more precision.

The main purpose of this paper is to implement the hydrodynamic approach in the context of perturbed non-equilibrium one-dimensional conformal field theory (Neq-CFT), and show that it emerges from the quantum description. 

Recall that Neq-CFT has been introduced to study out-of-equilibrium quantum phenomena using the extensive toolbox of conformal field theory (CFT) combined with an $S$-matrix approach \cite{BD11-12}. It aims at describing non-perturbatively, and beyond the linear response theory, the low-energy sector of gapless one-dimensional quantum systems driven far from equilibrium. The driving is obtained by unitary evolution of independently thermalized subsystems, a setup considered in other contexts before \cite{caro,rub,spo}. Exact results in one-dimensional Neq-CFT \cite{BD11-12} have been verified numerically in the Heisenberg chain \cite{Moore12}; other exact results along this line of thoughts have been obtained for free-fermionic quantum chains \cite{tas,ah,og,aschp,Luca_Viti,delucamv} and for higher-dimensional free models \cite{collura2,doyonKG}, and conjectures proposed for certain integrable systems \cite{doyonint,DeLucaVitiXXZ}. In higher-dimensional non-equilibrium CFT, a hydrodynamic approach was used within this setup in \cite{CKY,Nat-Phys}, leading to conjectured analytic formulas extending some of the results of \cite{BD11-12} and numerically verified within the Gauge/Gravity duality \cite{AY}. As explained in \cite{Nat-Phys}, the lack of integrability plays an important role for the hydrodynamic argument, and, combined with other ideas, the hydrodynamic data give the exact quantum density matrix of the steady state.

The CFT description of gapless systems is never exact but only asymptotically exact at low energy \cite{lowCFT}. The asymptotic approach to the low energy effective dynamics is controlled by the so-called irrelevant operators -- those which, by definition, do not influence the low-energy effective dynamics at equilibrium. Perturbations of CFTs by irrelevant operators are then important, especially far from equilibrium, in order to understand low-energy numerical \cite{Moore12} and experimental  \cite{Exp} results for non-equilibrium steady states and their universality \cite{conductance, Mintchev}, driven systems \cite{irrel-out} and quantum quenches \cite{Cardy-London}. In this paper we study Neq-CFT with a particular irrelevant perturbation related to effects of band curvature; as this breaks integrability, hydrodynamic ideas are expected to be fruitful in this situation.

\section{A hydrodynamic approach}

Before embarking into the hydrodynamic approach, let us recall some facts about Neq-CFTs \cite{BD11-12}. In the simplest case, one generates an out-of-equilibrium state in a CFT by first preparing two copies of this CFT on a semi-infinite half line each at different temperatures $T_l=\beta^{-1}_l$ and $T_r=\beta^{-1}_r$, and then gluing them at some initial time $t=0$ through a contact point, say the origin $x=0$. It was shown that a non-equilibrium steady state emerges on any interval $[-L,L]$ for all times $t>L/v_F$ where $v_F$ is the Fermi velocity; beyond the interval, the system remains in its thermal state. That is, two shock waves are associated to this energy flow, which are sharp within the CFT description and which propagate at the Fermi velocity, along the two branches of the light-cone. The interior region of the light-cone is in a non-equilibrium state carrying a mean energy current $J_E=\frac{c\pi}{12}(T^2_l-T_r^2)$ with $c$ the central of charge of the CFT, whereas the regions at the left and at the right of the light-cone remain at thermal equilibrium at temperature $T_l$ and $T_r$ respectively. Since there are no fixed reservoirs in this setup, in general local energy gradients decrease as time evolves, so that the emerging steady state must be controlled by ballistic transport. A similar framework for generating non-equilibrium steady states was first studied in the context of electronic transport \cite{caro,rub} and in harmonic chains \cite{spo} and originally named the partitioning approach.

Within the CFT approximation, the Hamiltonian is $H_{\rm CFT}=\int \dd x\,(T(x)+\bar T(x))$ where $T(x)$ and $\bar T(x)$ are the two chiral components of the stress tensor \cite{CFT-DiF}. They have scaling dimension $2$, so that the Hamiltonian has the correct dimension $1$ \footnote{By convention, we normalize the Fermi velocity to $v_F=1$. We set $k_B=1$ and $\hbar=1$.}. The most natural irrelevant operators are bilinear in the stress tensor components, that is $T\bar T$, and $T^2$ or $\bar T^2$.  For simplicity, we shall restrict ourselves to the $T \bar T$ perturbation, associated to curvature effects in the dispersion relation of the underlying gapless microscopic model.  The perturbed Hamiltonian we shall consider is thus
\begin{equation}\label{eq:hamilton}
 H= \int \dd x\, (T(x)+\bar T(x)) + g \int \dd x\, T(x)\bar T(x).
 \end{equation}
At equilibrium this perturbation is irrelevant because the operator $T\bar T$ has dimension $4$ and hence $g$ scales like $a^2$ with $a$ the short distance (UV) scale of the microscopic model.

In the following sub-sections we use a hydrodynamic approach to study the effects of the $T\bar T$ perturbation on the non equilibrium steady state, on the approach to the steady regime, and on the structure of the associated shock waves, and we show by a direct perturbative computation that these hydrodynamic considerations are correct.

\subsection{Generalities: hydrodynamic description near criticality}

The hydrodynamic approach is rooted in two principles \cite{ref-book-standard}: the first consists in assuming that there is some kind of local thermalization and the second that the equations of motion follow from conservation laws.

Local thermalization assumes that at every point $x$ and time $t$, the state may be described, locally in a neighborhood $\mathfrak{N}_{x,t}$, by an effective density matrix $\rho_{\rm local}$ of the form of an equilibrium density matrix, but involving all available local conserved charges of the theory. Near criticality in non-integrable models, omitting internal symmetries as we are only considering energy transport, the only charges that may describe transport are the hamiltonian $H$ and the momentum $P$, associated to invariance under time and space translations. In non-integrable systems there are no other local conserved charges. With $h$ and $p$ the hamiltonian and momentum densities respectively, $\rho_{\rm local}$ is of the form
\begin{eqnarray} \label{eq:rholocal}
\rho_{\rm local} \propto \exp\Big( - \beta_h(x,t)\int_{\mathfrak{N}_{x,t}} \hskip -0.4 truecm \dd x'\, h (x')+\beta_p(x,t)\int_{\mathfrak{N}_{x,t}}  \hskip -0.4 truecm \dd x'\, p (x') \Big)
\end{eqnarray}
with $x,t$-dependent parameters (local potentials) $\beta_h$ and $\beta_p$. This implies that the expectations of local observables can be computed as in a generalized thermal state of the form $e^{-\beta_h H + \beta_p P}$ (generalized by the presence of the momentum operator $P$), but with local effective inverse temperature $\beta_h$ and momentum potential $\beta_p$ that depend on space and time.

In quantum chain models, there is of course no continuous space translation, hence no conserved momentum operator. Continuous space translation only emerges near criticality, where low-energy excitations are supported on such large scales that the underlying lattice structure is not important. Without a conserved momentum operator, the density matrix is locally thermal, hence it cannot describe energy transport\footnote{With integrability, this problem can be circumvented by the presence of appropriate non-trivial conserved charges.}. It is the presence of a conserved momentum near criticality that allows, within this hydrodynamic description, for a possibly nonzero energy current to develop at large times in the partitioning approach. This agrees with the intuition according to which only ballistic currents may develop in this approach, while the lattice structure introduces diffusion away from criticality.

The choice of the local observables that should be described by local thermalization is important. For instance, in a thermal state, averages of derivatives of local observables are zero by translation invariance in space and time. On the other hand, in a locally thermalized state the derivative of the average of an observable is generically nonzero. Hence in a locally thermalized state, the choice of putting the derivative inside the average or outside matters, and this is guided by the physical scale on which the derivative is to be taken: either at the microscopic level (inside a hydrodynamic cell, where quantities are effectively constant), or at the hydrodynamic level (there are variations from one hydrodynamic cell to the next). In the hydrodynamic description of QFT one assumes that conserved densities and currents are locally thermalized, and that hydrodynamic equations are determined by the conservation laws.

In the present situation we have two local conservation laws, those of time and space translation invariance:
\begin{eqnarray} \label{eq:conserve}
	\p_t h + \p_x j = 0,\quad \p_t p + \p_x k = 0.
\end{eqnarray}
Within the (pure) hydrodynamic approximation, expectations of the densities $h$ and $p$ and the currents $j$ and $k$ are evaluated by local thermalization: they are first evaluated in generalized thermal states, then the local potentials are made dependent on space and time. Hence they become functions of the local potentials $\beta_h(x,t)$ and $\beta_p(x,t)$. Let us denote these expectations by $\hh$, $\pp$ for the densities and $\jj$ and $\uu$ for the currents. From eq.(\ref{eq:conserve}) we get two equations
\begin{eqnarray} \label{eq:hydro}
 \p_t \hh + \p_x \jj=0,\quad \p_t \pp + \p_x \uu =0 .
 \end{eqnarray}
These are non-linear differential equations for $\beta_h$ and $\beta_p$, which are the hydrodynamic equations. Notice that we have two equations for two functions with given initial condition, so that in principle the problem is reduced to solving non-linear PDEs.

In thermal equilibrium, the energy current and momentum density are zero by time-reversal and space-reversal symmetry. Since there is only one parameter (the temperature), there is an equation of state $\uu = F(\hh)$ relating the pressure $\uu$ to the energy density $\hh$. In the generalized state, the equations of state may be generalized to $\uu = F(\hh,\pp)$, and $\jj = G(\hh,\pp)$. With the generalized equations of state, the hydrodynamic problem is recast into equations for local energy and momentum densities, instead of equations for $\beta_h$ and $\beta_p$

Other parametrizations of the hydrodynamic problem are possible,  and one often uses ``fluid velocities''. In some situations, it is possible to use a parametrization that explicitly separates ``dynamical'' degrees of freedom from ``thermal'' ones. Assume that there is an additional dynamical symmetry of boost: a space-time transformation operator $B$ that transforms energy into momentum, $[B,H] = P$, and that preserves the space spanned by $H$ and $P$. Two immediate examples are the relativistic and the galilean boosts, $[B,P]=H$ (with Casimir $H^2-P^2$) and $[B,P]=m{\bf 1}$ (with Casimir $H-P^2/(2m)$), respectively. The former applies to critical points with dynamical exponent $z=1$, the latter with $z=2$ (see for instance \cite{Joe-Ben} for a hydrodynamic study of non-equilibrium states in this context). Using the boost operator, we can always write $\beta_hH - \beta_p P = \beta_{\rm rest}\, e^{-\theta B} H e^{\theta B}$. The choice of $B$ is not unique, and  for every space-time point $x,t$ we have a boost $B_{x,t}$ that keeps $x,t$ invariant. The local density matrix can be written as a local boost of a local thermal state,
\begin{eqnarray} \label{eq:rholocal}
\rho_{\rm local} \propto \exp\Big( - \beta_{\rm rest}(x,t)\int_{\mathfrak{N}_{x,t}} \hskip -0.4 truecm \dd x'\, e^{-\theta(x,t) B_{x,t}} h (x') e^{\theta(x,t) B_{x,t}}\Big).
\end{eqnarray}
The local boost parameter $\theta(x,t)$ is the dynamical part, and the  local rest-frame inverse temperature $\beta_{\rm rest}(x,t)$ is the thermal part. This hydrodynamic parametrization was used in \cite{Nat-Phys} in order to give a characterization of higher-dimensional non-equilibrium CFT and describe the steady state as a boosted thermal state.

It turns out that the requirement that the momentum density be equal to the energy current up to, possibly, an even derivative term, $p=j + \mu\,\p_x^{2\ell}j$, is essentially enough to imply relativistic invariance. Further, this implies that, in a thermal state,
\beq\label{relthermo}
	T\frc{d}{dT} \uu = \hh + \uu,
\eeq
and that, given the thermal equation of state $\uu = F(\hh)$, all generalized thermal averages are explicitly expressed as functions of $\theta$ and $\hh$, see eq.\eqref{p1} in Appendix \ref{apprel}. These facts, which we show within the general context of local many-body systems in Appendix \ref{apprel}, can be useful to simplify the description of the hydrodynamic problem.

Finally, the hydrodynamic approximation discussed above is the ``pure'' hydrodynamics. It can be made more accurate by adding {\rm derivative corrections}, including viscosity terms. These are corrections, in the expressions for the densities and currents, involving higher derivatives of the basic quantities (for instance corrections to the equations of state  involving derivatives of energy and momentum densities), and are associated with shrinking the hydrodynamic cell in order to take into account more and more of the microscopic variations.

We will make all these concepts more precise in the example of the perturbed CFT studied in the present paper.

\subsection{The exact first-order hydrodynamic description of $T\b T$ perturbed CFT}

In the present setting of perturbed conformal field theory, the hamiltonian density could be taken as $T(x) + \bar T(x) + g T(x) \bar T(x)$, and the momentum density $p(x) = T(x) - \bar T(x)$. Of course, these are defined up to total derivatives of local densities, and for our purposes it will be convenient to add the term $-gc/(24\pi)\p_x^2(T(x)+\b T(x))$ to the above energy density. The equations of motion are then determined by the Hamiltonian (\ref{eq:hamilton}), via $\p_t O= i[H,O]$ for any operator $O$. The basic commutation relations for the stress-energy tensor are 
\[ -i[T(x),T(y)]=-(T(x)+T(y))\,\delta'(x-y) +\frac{c}{24\pi}\delta^{'''}(x-y),\]
with $c$ the central charge of the CFT \cite{CFT-DiF}, and similarly for the $\bar T$ commutation relation except for the change of $i \to -i$. The $T$ and $\bar T$ components commute. Hence, we have
\begin{eqnarray} 
h(x) &=& T(x) + \bar T(x) - \frc{gc}{24\pi} (T''(x)+ \b T''(x)) + g T (x)\bar T(x)\n \label{eq:defhj1}
 j(x) &=& T(x)-\b T(x) + O(g^2)
\end{eqnarray}
for the energy density and current, and
\begin{eqnarray} 
p(x) &=& T(x) - \bar T(x) \n
 k(x) &=& T(x) + \bar T(x) - \frc{gc}{24\pi} (T''(x)+ \b T''(x)) + 3g T(x) \b T(x)
 + O(g^2)\label{eq:defhj2}
\end{eqnarray}
for the momentum density and current. Here and below primes denote space derivatives.

Let us briefly analyze the case $g=0$. In this case, the local density matrix is  \begin{eqnarray} \label{eq:rholocal}
\rho_{\rm local} \propto \exp\Big( - \beta(x,t)\int_{\mathfrak{N}_{x,t}} \hskip -0.4 truecm \dd x'\, T (x')-\bar \beta(x,t)\int_{\mathfrak{N}_{x,t}}  \hskip -0.4 truecm \dd x'\,\bar T (x') \Big)
\end{eqnarray}
with $x,t$-dependent parameters $\beta=\beta_h-\beta_p$ and $\bar \beta=\beta_h+\beta_p$, and we have $j=p=T-\b T$ and $k=h= T+\b T$. Hence the averages of $T$ and $\b T$ can be used as hydrodynamic variables, and the expectations of any local observables, say products or powers of $T$ and $\bar T$ at some point $x$ and time $t$, can be computed, within the hydrodynamic approximation, as in equilibrium CFT but with local effective ``temperatures'' $\beta(x,t)$ and $\bar \beta(x,t)$. In particular, there is chiral factorization: right-movers and left-movers are independently thermalized. At $g=0$ the CFT enjoys relativistic invariance, with boost operator $B=-i\int \dd x\,xh(x)$ satisfying $[B,H] = P$ and $[B,P]=H$, and the above can be interpreted as a boosted density matrix with rest-frame inverse temperature $\beta_{\rm rest} = \sqrt{\beta\b\beta}$ and boost parameter given by $\tanh\theta = (\b\beta-\beta)/(\b\beta+\beta)$. The thermal average at temperature $\tau$ is \cite{Cardy, Affleck}
\beq\label{zeroth}
	\bra T\ket = \bra \b T\ket = \frc{c\pi}{12} \tau^2,
\eeq
and with the initial condition where left and right halves are independently thermalized at temperatures $T_l$ and $T_r$ respectively, the hydrodynamic problem gives $\beta(x,t) = T_l^{-1} \Theta(x+t)$ and $\b\beta(x,t) = T_r^{-1} \Theta(-x+t)$. As was shown in \cite{BD11-12,Nat-Phys}, this hydrodynamic description is in fact {\em exact}, and does not necessitate any derivative corrections: it is a direct consequence of chiral factorization at the level of the quantum equations of motion. This solution represents sharp shocks emanating from the contact point and propagating at the speed of light in opposite directions, between which the steady states lies.

Let us now consider the first order correction in $g$. We first use the facts that, in a generalized thermal state, there is translation invariance, $\bra T''\ket = \bra \b T''\ket=0$, and that to leading order in $g$ there is chiral factorization $g\bra T\b T\ket = g\bra T \ket \bra \b T\ket + O(g^2)$. We can then parametrize averages using the two quantities $w(x,t) = \bra T(x,t)\ket$ and $\b w(x,t)= \bra \b T(x,t)\ket$ evaluated within the pure hydrodynamic approximation using $\rho_{\rm local}$, and we get
\beq\label{noder}
\begin{aligned}
& \hh = w + \bar w + g\, w \bar w + O(g^2),\quad
\jj = w-\b w  + O(g^2),\\
& \pp = w - \bar w,\quad
\uu = w + \bar w + 3g\, w \b w +O(g^2).
\end{aligned}
\eeq
Recall that the equations of motion are $\p_t \hh + \p_x \jj=0$ and $\p_t \pp + \p_x \uu =0$. These are two equations for the two unknowns $w$ and $\bar w$, chosen as a parametrization of the two unknown local potentials $\beta_h$ and $\beta_p$.

This description is {\em not} exact at order $g$: there are derivative corrections. Fortunately, in the present case it is possible to obtain the full, exact derivative corrections at order $g$: they are a direct consequence of the quantum equations of motion. It turns out that these derivative corrections are ``trivial'': they can be entirely absorbed into appropriate definitions of basic hydrodynamic quantities, in such a way that the  hydrodynamic equations remain unchanged. Writing the conservation laws for quantum averages using \eqref{eq:defhj1} and \eqref{eq:defhj2} without assuming local thermalization, using only chiral factorization at $g=0$, the definitions
\beq \label{eq:defw}
\begin{aligned}
 w(x,t) = \bra T(x,t)\ket - \frc{gc}{48\pi}\lt(\bra T''(x,t)\ket + \bra \b T''(x,t)\ket\rt)\\
 \b w(x,t)= \bra \b T(x,t)\ket - \frc{gc}{48\pi}\lt(\bra T''(x,t)\ket + \bra \b T''(x,t)\ket\rt)
 \end{aligned}
\end{equation}
give rise to the pure-hydrodynamic equations for the quantities \eqref{noder}. That is, in terms of these exact quantum averages, the conservation equations for the densities and currents \eqref{noder} are a consequence of the quantum equations of motion, hence are {\em exact} for all $x$ and $t$ to order $g$. All derivative corrections have been absorbed into the definitions \eqref{eq:defw} and the choice of the energy density in \eqref{eq:defhj1}.

The initial conditions are those reflecting that initially the left and right parts are independently thermalized at temperatures $T_l$ and $T_r$ respectively. At thermal equilibrium with temperature $\tau$, the average momentum density is zero (and averages are space-time independent), hence $w=\b w=:w_{\rm th}(\tau)$. By dimensional analysis
\beq\label{wth}
	w_{\rm th}(\tau) = \frc{c\pi}{12} \tau^2 f(g\tau^{2})
\eeq
for some function $f$ with $f(0)=1$.  We expect $f$ to have an expansion in positive integer powers of $g$, and the normalization of the zeroth order \eqref{zeroth} gives
\beq\label{f}
	f(u) = 1 + au + O(u^2).
\eeq
Hence, at initial time,
\beqa
	&& w=\bar w=\frac{c\pi}{12}T^2_l\lt(1+agT_l^2 + O(g^2)\rt)\quad \mbox{on the left ($x<0$)}\n
	&& w=\bar w=\frac{c\pi}{12}T^2_r\lt(1+agT_r^2+O(g^2)\rt)\quad \mbox{on the right ($x>0$)}\label{init}
\eeqa
with an abrupt crossover in the neighborhood of the contact point (that is, close to the origin). Below we compute the constant $a$ by using relativistic invariance, giving
\beq\label{a}
	a = -\frc{c \pi}6.
\eeq
This fully fixes the hydrodynamic problem (with all derivative corrections) at order $g$.
\medskip

\begin{rema}
Given any solution $w(x,t)$ and $\b w(x,t)$, the quantum averages $\bra T(x,t)\ket$ and $\bra \b T(x,t)\ket$, obtained by solving \eqref{eq:defw}, are naively not uniquely fixed:
\[\begin{aligned}
	\bra T(x,t)\ket &= w(x,t) + \frc{gc}{48\pi}\lt(w''(x,t)+\b w''(x,t)\rt)
	+ \omega(x,t)\\
	\bra \b T(x,t)\ket &= \b w(x,t) + \frc{gc}{48\pi}\lt(w''(x,t)+\b w''(x,t)\rt)
	+  \omega(x,t)
	\end{aligned}
\]
where $\omega(x,t)$ is solution of the equation $\omega -\frac{gc}{24\pi}\,\omega''=O(g^2)$ whose kernel is of the form $A_+ e^{x/\sqrt{gc/(24\pi)}} + A_- e^{-x/\sqrt{gc/(24\pi)}}$. However, this kernel is non-perturbative in $g$, hence beyond the present order-$g$ calculation.
\end{rema}

\subsection{Equation of state and relativistic structure at order $g$}

The thermal equation of state relating the pressure $\uu$ to the energy density $\hh$ can be obtained from \eqref{noder} by using the fact that $w=\b w$ in a thermal state:
\beq\label{eos}
	\uu = F(\hh) = \hh +\frc g2\hh^2 + O(g^2).
\eeq
The generalized equations of state can be obtained similarly,
\beq\label{geos}
	\uu =  \hh +\frc g2(\hh^2 - \pp^2) + O(g^2),\quad \jj = \pp + O(g^2).
\eeq
Associated to the equilibrium equation of state \eqref{eos} is the sound velocity (at temperature $\tau$)
\beq\label{sound}
	v_{\rm s} = \sqrt{F'(\hh)} = 1+\frc{g}2\hh + O(g^2) = 1 + \frc{gc\pi}{12} \tau^2 + O(g^2)
\eeq
Remark from \eqref{noder} that to order $g$, we have $\jj = \pp$. This equality implies propagation of small disturbances of energy, momentum and pressure densities near their equilibrium values occurs through sound-like waves with this sound velocity (see for instance the arguments presented in \cite{Levitov,D_super}). We will see below that the sound velocities also play a role far from equilibrium at order $g$.

Further, observe from \eqref{eq:defhj1} and \eqref{eq:defhj2} that the equality $j = p$ holds at the operator level at order $g$. This implies that relativistic invariance emerges (see Appendix \ref{apprel} for the general theory). Indeed one can verify explicitly that the boost operator $B = -i\int \dd x\,xh(x)$, with $h(x)$ the perturbed energy density \eqref{eq:defhj1}, satisfies $[B,H] = P +O(g^2)$ and $[B,P]=H$. Therefore, the averages in the generalized thermal state are expressed in terms of relativistically boosted observables in a thermal state. Using the result \eqref{p1} along with \eqref{eos}, we may express them as
\beqa
	\hh &=& \cosh 2\theta\,\hh_{\rm rest}
	+ \frc{g}4 (\cosh2\theta-1)\,\hh_{\rm rest}^2 + O(g^2)\n
	\jj=\pp &=& \sinh2\theta\,\lt(\hh_{\rm rest}
	+ \frc g4\, \hh_{\rm rest}^2\rt) +O(g^2)\n
	\uu &=& \cosh 2\theta\,\hh_{\rm rest}
	+ \frc{g}4 (\cosh2\theta+1)\,\hh_{\rm rest}^2 +O(g^2).
	\label{relg}
\eeqa
where $\hh_{\rm rest}$ is the rest-frame thermal energy density $\hh_{\rm rest} = \hh_{\rm th}(T_{\rm rest})$, deduced from 
\eqref{noder}, \eqref{wth} and \eqref{f}, given by $\hh_{\rm th}(\tau) = \frc{c\pi}{6} \tau^2\lt(1+g\tau^2\lt(a + \frc{c\pi}{24}\rt) + O(g^2)\rt)$. The relation \eqref{relthermo} can be combined with the thermal equation of state \eqref{eos} and the form \eqref{hth} of the thermal energy density. A simple calculation shows that this fixes $a$ to \eqref{a}, and thus
\beq\label{hth}
	\hh_{\rm th}(\tau) = \frc{c\pi}{6} \tau^2\lt(1-\frc{gc\pi}{8}\tau^2 + O(g^2)\rt).
\eeq
\medskip

\begin{rema} \label{rema2} Formally extending the above thermodynamic equations to large values of $|g|$ one finds that, in some respects, $g<0$ appears to be more natural. For instance, the energy density \eqref{hth} has an unphysical maximum, as a function of temperature, if $g>0$. Further, if $g>0$, the sound velocity \eqref{sound} is greater than the Lorentz speed of light (which is set to unity) associated to the (order-$g$) relativistic invariance. However, these facts do not lead to perturbative inconsistencies, and none of the perturbative calculations below are affected by the sign of $g$. The temperature at which the energy density is maximal is nonperturbative (of order $1/g$); and a Lorentz transformation from a point in the space-time region lying between the light cone and the sound cone to a point at a time $t=0$ would involve a very large boost parameter (non-perturbative in $g$), thus precluding, in perturbative considerations, the conclusion of nonzero equal-time correlations (breaking of causality). In fact, the Lieb-Robinson velocity $v_{LR}$ can be shown perturbatively to be greater than $1+Ag$ for any constant $A>0$, and thus greater than the sound velocity no matter the sign of $g$. Below we keep $g$ arbitrary, and all conclusions stay valid perturbatively in $g$ independently of its sign.
\end{rema}

\subsection{Light cone effects, shocks and steady states}

In this section, we solve the hydrodynamic problem (\ref{eq:hydro}) associated to the $T\bar T$ perturbation, using the parametrization in terms of the functions $w$ and $\b w$ in Eqs. \eqref{noder}. We decipher how the $T\bar T$ perturbation modifies the shock propagation, the structure of the shocks as well as the approach to the steady states and the steady state itself. We shall work to first order in $g$ only.

\subsubsection{Shock wave assumption}\label{sssectshocks}

The simplest analysis of the hydrodynamic problem is that which assumes that the picture of shocks emanating from the contact point, which holds exactly at $g=0$, is accurate even with $g\neq0$ at large scales. This picture was used in \cite{CKY,Nat-Phys} in order to generalize results to higher-dimensional CFT. More precisely, we make the assumption that, with exponential accuracy, averages of local observables as functions of the space-time point $(x,t)$ take steady and uniform values whenever $|x+v_lt| \gg \xi_l(t)$ and $|x-v_rt| \gg \xi_r(t)$, for some $\xi_{l,r}(t)$ bounded from above by a power law $\propto t^a$ for all $t$ with some exponent $a<1$. This represents shocks at speeds $v_l>0$ and $v_r>0$, propagating towards the left and the right respectively, with time-dependent widths $\xi_l(t)$ and $\xi_r(t)$, respectively, which grow sublinearly at large times. By the initial conditions, two the left (right) of both shocks, averages are evaluated in the initial left (right) reservoir, and the steady state lies between both shocks, described by a generalized thermal state.

It turns out that this shock assumption, along with the knowledge of the averages in the generalized thermal states, is sufficient to fix both the shock velocities and the steady-state parameters. It does not, however, give any information on the approach to the steady state and on the structure of the shocks themselves.

Consider the conservation laws \eqref{eq:conserve} integrated along a rectangular contour crossing a shock and whose diagonal goes from $(x_1,t_1)$ to $(x_2,t_2)$. That is,
\[
	\int_{x_1}^{x_2} \dd x\, (h(x,t_2) - h(x,t_1))
	+ \int_{t_1}^{t_2} \dd t\, (j(x_2,t)-j(x_1,t)) = 0,
\]
with $(x_1,t_1)$ and $(x_2,t_2)$ both lying inside the shock, and similarly for $p$ and $k$. These equations can be written for the left and the right shocks. We can choose the rectangle such that $|x_2-x_1|=v_{l,r}|t_2-t_1|$ with $|x_2-x_1|$ much larger than the width of the shock. Using the assumptions that the densities and the currents are asymptotically uniform and steady away from the shocks, that the shocks are of extent that grows sublinearly in time, and that the current vanishes outside the light-cone, this yields two sets of two equations, one for the left shock the other for the right shock,
\beq\label{shockseq}
	v_l(\hh_l - \hh_{s}) = \jj_{s},\quad
	v_l \pp_{s} = \uu_{l} - \uu_{s},\quad
	v_r(\hh_{s}-\hh_{r}) = \jj_{s},\quad
	v_r \pp_{s} = \uu_{s} - \uu_{r},
\eeq
where indices $l,\,r,\,s$ indicate the averages in the left reservoir, right reservoir and in the steady-state region, respectively. 
Adding them allows to eliminate $\hh_s$ and $\uu_s$, keeping only the densities $\hh_{l,r}$ and the pressures $\uu_{l,r}$ evaluated in the asymptotic region at equilibrium,
\begin{equation}\label{eexx}
 (v_l+v_r)\, \pp_s= \uu_l-\uu_r,\quad (v_l^{-1}+v_r^{-1})\, \jj_s = \hh_l-\hh_r.
 \end{equation}
Combining them, we find the two equations
\begin{equation}\label{ee}
	(\hh_l-\hh_s)(\uu_{l}-\uu_{s}) =
	(\hh_r-\hh_s)(\uu_r-\uu_s)
	= \jj_{s}\pp_{s}.
\end{equation}
These are two equations for two steady-state unknowns. 
The steady-state unknowns can be taken as $w_s$ and $\b w_s$ using the parametrization \eqref{noder}, the boost and rest-frame temperature $\theta$ and $T_{\rm rest}$ using the relativistic parametrization \eqref{relg}, or simply the steady-state energy density and current $\hh_s$ and $\jj_s$ using the generalized equations of state \eqref{geos}. A calculation up to order $g$ gives, within the latter parametrization,
\beqa
	\hh_s &=& \frc{\hh_l + \hh_r}2 + \frc g8 (\hh_l-\hh_r)^2+ O(g^2)\n
	\jj_s &=&
	\frc{\hh_l-\hh_r}2 + \frc{g}8
	(\hh_l^2-\hh_r^2) +O(g^2).\label{twoshockssol}
\eeqa
The thermal averages $\hh_{l,r}$ can be evaluated in terms of the temperatures $T_{l,r}$ using \eqref{hth}, and in particular we obtain
\beq
	\jj_s =
	\frc{c\pi}{12} T_l^2\lt(1-\frc{gc\pi}{12} T_l^2 \rt)
	-
	\frc{c\pi}{12} T_r^2\lt(1-\frc{gc\pi}{12} T_r^2 \rt)
	+ O(g^2).
\eeq
The shock velocities can now be evaluated through
\beq
	v_{l,r} =
	\lt|\frc{\jj_s}{\hh_{l,r} - \hh_s}\rt|
	= 1+\frc{gc\pi}{12} T_{l,r}^2 + O(g^2).
\eeq

We make several observations.
\bi
\item First, to this order in $g$, the steady-state current is still a difference of a function of the left-reservoir temperature minus the same function of the right-reservoir temperature, $\jj_s = J(T_l) - J(T_r)$. In the unperturbed case this had important echoes on the full counting statistics, and in general this implies that the non-equilibrium current can be obtained purely from the linear-response conductivity $G(\tau) = d\jj_s/dT_l\big|_{T_l=T_r=\tau} =  dJ(\tau)/d\tau$ as $\jj_s = \int_{T_r}^{T_l}d\tau\,G(\tau)$ \cite{Moore12}. 
\item Second, the shock velocities $v_l$ and $v_r$ are exactly the sound velocities \eqref{sound} of the left and right reservoir, respectively. That is, it is the linear sound velocities of the reservoirs that control the speed of the shocks describing the far-from-equilibrium steady state.
\item Third, parametrizing the reservoirs using $w$ (with the thermal result \eqref{wth}, \eqref{f}, \eqref{a}) we note that the current takes the simple form
\beq\label{jw}
	\jj_s = v_lw_l - v_rw_r  + O(g^2).
\eeq
This can naturally be interpreted as the ballistic transport of energy by independent left/right movers, with chiral energy densities $w_{l,r}$ and velocities $v_{l,r}$ that are solely determined by the reservoirs' temperatures. One may understand this as a generalization of the case $g=0$, where $w$ and $\b w$ are right- and left-moving quantities at the speed 1: to order $g$, $w$ and $\b w$ may still be seen as independent right- and left-moving energy densities, but their velocity is ``dressed'' into the sound velocity by the reservoirs' state. This is also an extension of the picture that holds near equilibrium. With $T_l\approx T_r$, there is linear wave propagation with the sound velocity $v$. The natural independent right- and left-moving energy densities are the combinations of space-time dependent small variations (near the equilibrium values) $\delta\varep_\pm = (\delta\hh\pm v^{-1}\delta\jj)/2$, and the current is $v(\delta\varep_+-\delta\varep_-)$.
\item Fourth, near equilibrium (i.e. with $T_l\sim T_r$), linear wave propagation implies that \cite{D_super} $\jj_s \sim (\uu_l - \uu_r)/(2v)$ where $v$ is the equilibrium sound velocity. One can indeed check that this agrees with \eqref{jw}, using $d(vw)/d\tau = (1/2v)\,d\uu/d\tau$ (where $v,\,w,\,\uu$ are at equilibrium with temperature $\tau$).
\item Fifth, one can verify that the inequality $\jj_s > (\uu_l - \uu_r)/(2v_{LR})$ \cite{D_super} is verified, where $v_{LR}$ is the Lieb-Robinson velocity (see Remark \ref{rema2} concerning $v_{LR}$).
\item Finally, we quote the steady-state rest-frame temperature and boost velocity from its relativistic parametrization:
\beq
	T_{\rm rest} = \sqrt{T_lT_r}\lt(1-\frc{gc\pi}{48}(T_l-T_r)^2 + O(g^2)\rt),\quad
	\tanh\theta = \frc{T_l-T_r}{T_l+T_r}\lt(1-\frc{gc\pi}{12}T_lT_r
	+O(g^2)\rt).
\eeq
In particular, this means that in the steady state, the potential $\beta_h$ associated with the hamiltonian in the density matrix as in \eqref{eq:rholocal} takes the simple form
\beq
	\beta_h = \frc{\beta_l+\beta_r}2.
\eeq
\ei

\subsubsection{Solution to the hydrodynamic problem: appearance of shocks}

In the previous paragraph, we have made the assumption that two shocks emanate from the connection space-time point. In the present paragraph, we provide the solution to the full order-$g$ hydrodynamic problem. The main observations from the calculation below are as follows:
\bi
\item The problem does not have a unique solution. The space of solution includes the two-shock solution above, but also contains solutions with additional ``remnant'' shocks at $x=\pm t$.
\item Nevertheless, the steady state is unique: every solution gives rise to the same state in the central region (this state is in agreement with that found from the two-shock assumption).
\ei

Let us discuss briefly the first point above. The non-uniqueness of the weak solutions to the Riemann problem (the step initial conditions) for pure hydrodynamic equations has been observed before \cite{Nat-Phys}. In standard hydrodynamic problems, it is usually possible to lift this non-uniqueness by considering the entropy current or any other inequality condition \cite{Bressan}. With the requirement that the entropy cannot decrease at the shocks (a local implementation of the $2^{\rm nd}$ law of thermodynamics), this may forbid certain shocks and lead to rarefaction waves (transition regions of extent growing linearly with $t$). Positivity of local entropy production occurs with viscosity terms, and weak solutions to pure hydrodynamics should be seen as emerging, at large scales, from viscous hydrodynamics; thus positive local entropy production at the shocks becomes a condition for selecting weak pure hydrodynamics solutions. In the present case, higher-derivative corrections are completely absorbed into redefinitions, hence at order $g$ there is no positivity condition that can fix the solution. We hope to come back to this problem in the future by analyzing higher orders in $g$.

Let us proceed with the calculation. The full order-$g$ hydrodynamic problem is given, in terms of the parametrization $w$, $\b w$ as defined in \eqref{eq:defw}, by the conservation equations \eqref{eq:hydro} with \eqref{noder}, along with the initial conditions \eqref{init}.

At $g=0$, denoting $w=w_0$, these equations simplify to $\p_xw_0 = -\p_t w_0$ and $\p_x \b w_0 = \p_t \b w_0$. With the initial conditions, the solution is immediately given by
\beq
	w_0 = \frc{c\pi}{12} \lt(T_l^2 \Theta(t-x) + T_r^2\Theta(x-t)\rt),\quad
	\b w_0 = \frc{c\pi}{12} \lt(T_l^2 \Theta(-t-x) + T_r^2\Theta(x+t)\rt).
\eeq
This indeed represents two sharp shock waves both at velocity 1.

Up to first order in $g$, we may solve the hydrodynamic problem by using this zeroth-order solution in order to simplify the bilinar terms proportional to $w\b w$ in \eqref{noder}. The main idea is that, at order $g$, we may replace $gw$ and $g\b w$ by $gw_0$ and $g\b w_0$ respectively. We first divide space-time (with positive times $t>0$) into three regions: the left (L) $x<t$, the center (C) $-t<x<t$ and the right (R) $x>t$. Within each of these regions, $w_0$ and $\b w_0$ are both constant, with $w_0 = \frc{c\pi}{12} T_l^2$ in the center and left, $w_0 =\frc{c\pi}{12} T_r^2$ in the right; and $\b w_0 = \frc{c\pi}{12} T_r^2$ in the center and right, $\b w_0 = \frc{c\pi}{12} T_l^2$ in the left. Hence we may write $g\p_t(w\b w) \approx gw_0 \p_t \b w + g\b w_0 \p_t w$ with $w_0$ and $\b w_0$ piecewise constant, and similarly for $g\p_x (w\b w)$. These transform the bilinear terms into linear terms with piecewise constant coefficients. The equations then become
\beq  \label{eq:linear}
\begin{aligned}
	(\p_x +(1+g\b w_0)\p_t)w- (\p_x-(1+g w_0)\p_t)\b w  &= 0 \\
	(\p_t+(1+3g\b w_0)\p_x) w -(\p_t -(1+3g w_0) \p_x) \b w  &= 0.
\end{aligned}
\eeq

Eqs. \eqref{eq:linear} are linear equations which we can solve, within each region, by looking for solutions of the form $e^{i(px-E_pt)}$. A straightforward calculation shows that these plane waves have dispersion relations with two branches:
\beq
	E_p = \lt\{\begin{aligned}
	&p(1+g\b w_0 + O(g^2))\\
	& -p(1+gw_0 + O(g^2)).
	\end{aligned}
	\rt.
\eeq
We observe that the dispersion relation, to order $g$, is still {\em linear}, and hence only changes the propagation velocities.
This means that the solutions for $w(x,t)$ and $\b w(x,t)$ have the form, up to $O(g^2)$ terms,
\beq\label{solw}
	w(x,t) = w^+(x-v^+t) + w^-(x+v^-t),\quad
	\b w(x,t) = \b w^+(x-v^+t) + \b w^-(x+v^-t)
\eeq
where $v^+ = 1+g\b w_0 $ and $v^- = 1+g w_0$. In the left ($L$), center ($C$) and right ($R$) regions we obtain the velocities:
\beq\label{vpm}
	v^\pm_L =v_C^-=  1 + \frc{gc\pi}{12} T_l^2,\quad 
	v_R^\pm = v^+_C = 1 + \frc{gc\pi}{12} T_r^2,
\eeq
and within these regions we denote the left- and right-moving waves by $w^\pm_{L,C,R}$ and $\b w^\pm_{L,C,R}$. Note that the propagation velocities \eqref{vpm} are the shock velocities $v_l$ and $v_r$ obtained in paragraph \ref{sssectshocks} from assuming the presence of two shocks.

For definiteness we assume that $g>0$, so that all velocities are greater than one; a similar calculation can be done for $g<0$, leading to the same conclusions. 

The functions $w^+$, $w^-$, $\b w^+$ and $\b w^-$ in \eqref{solw} are related to each other. This can be obtained by putting \eqref{solw} into any one of the two equations \eqref{eq:linear}, for instance the first one. Since the arguments of the functions are either $x-v^+t$ or $x+v^-t$ and the equation holds for all $x$ and $t$ (within a given region), this single equation leads to two relations, one for the right movers the other for the left movers. These two relations contain single-derivative terms with constant coefficients, as in \eqref{eq:linear}. Integrating, one obtains similar relations but without derivatives. The integration constants may be set to zero, which accounts for the ``gauge symmetry'' in \eqref{solw},  $w^\pm\mapsto w^\pm \pm A$ and $\b w^\pm\mapsto \b w^\pm \pm B$ for constants $A$ and $B$.
After these operations, we find that the relations are
\beq\label{relation}
	(1 \mp v^+v^\pm) w^\pm = (1\pm v^+v^\mp) \b w^\pm + O(g^2).
\eeq
Since the velocities are of the form $v^\pm = 1+O(g)$, we see that $\b w^+ = O(g)$ and $w^- = O(g)$. It is then convenient to work solely with $w^+$ and $\b w^-$ instead, and using \eqref{relation} and the form of $v^\pm$ we find that \eqref{solw} becomes
\beq\label{wbw}
	w = w^+ - gw_0 \b w^- + O(g^2),\quad
	\b w = \b w^- - g\b w_0 w^+  + O(g^2).
\eeq

The boundary conditions are conditions on the functions $w_{R,L}$ and $\b w_{R,L}$. On the right, we have $w_R(x,0) = \b w_R(x,0) = w_r$ for all $x>0$, and on the left we have $w_L(x,0) = \b w_L(x,0) = w_l$ for all $x<0$, where we recall that $w_{l,r}$ are the thermal values $w_{\rm th}(T_{l,r})$. Using the decomposition \eqref{wbw} and the velocities \eqref{vpm}, as well as the leading part of the thermal values $w_{l,r} = (c\pi/12) T_{l,r} + O(g)$, this gives
\beq\label{initcond}\begin{aligned}
	w_R^+(x) &= \b w_R^-(x) = w_r + g w_r^2 +O(g^2)\qquad (x>0)\\
	w_L^+(x) &= \b w_L^-(x) = w_l + g w_l^2 +O(g^2)\qquad (x<0).
	\end{aligned}
\eeq

The full solution to the problem is obtained by supplementing the boundary conditions with the continuity conditions between the regions, at the interfaces $x=\pm t$. The continuity conditions are obtained by considering the integrated version of the equations \eqref{eq:linear}, which, we recall, are the full order-$g$ equations and are equivalent to \eqref{eq:hydro}.
Consider first the jumps at the interface $x=t$ between the center and right regions. For a shock of velocity 1 between the center and right regions across the interface $x=t$, this is (using $\jj=\pp+O(g^2)$)
\[
	\hh_C-\hh_R = \jj_C - \jj_R = \uu_C - \uu_R.
\]
From \eqref{noder}, the first equation is equivalent to
\[
	\b w_R-\b w_C = \frc{g}2 (w_C - w_R) \b w_R + O(g^2),
\]
while the second is similar but with $g$ replaced by $3g$. Combining, one possible solution is $\b w_R = O(g)$; but it turns out that this is inconsistent with the boundary condition. Hence the only consistent solution is
\beq\label{bwcont}
	w_R-w_C = O(g),\quad \b w_R - \b w_C = O(g^2)\qquad
	\mbox{(at $x=t$).}
\eeq
That is, to order $g$ there is no jump in $\b w$ at that interface, but there may be an $O(g)$ jump in $w$. A similar calculation at the interface $x=-t$ gives
\beq\label{wcont}
	\b w_L - \b w_C = O(g),\quad w_L - w_C = O(g^2)
	\qquad
	\mbox{(at $x=-t$).}
\eeq

Let us now combine the continuity condition at the interface $x=t$ with the boundary conditions. Using the decomposition \eqref{wbw} and the velocities \eqref{vpm}, the second relation of the continuity condition \eqref{bwcont} is
\[
	\b w_C^-((1+v_l)t) - g w_r w_C^+((1-v_r)t)
	= \b w_R^-((1+v_r)t) - gw_r w_R^+((1-v_r)t)  + O(g^2)
\]
while the first is $w_C^+((1-v_r)t) = w_R^+((1-v_r)t) + O(g)$, thus leading to
\beq
	\b w_C^-((1+v_l)t) = \b w_R^-((1+v_r)t) + O(g^2).
\eeq
With the initial conditions \eqref{initcond} we then find
\beq\label{bwc}
	\b w_C^-(x) = w_r + g w_r^2 + O(g^2) \quad (x>0).
\eeq
A similar calculation combining the initial conditions with the continuity conditions \eqref{wcont} at the interface $x=-t$ gives
\beq\label{wc}
	w_C^+(x) = w_l + g w_l^2 +O(g^2) \quad (x<0).
\eeq
These then imply, using the above assumption that the velocities are greater than 1,
\beq\label{restRL}
	w_R^+(x) = w_l +O(g) \quad (x<0),\qquad
	\b w_L^-(x) = w_r + O(g) \quad (x>0).
\eeq

Equations \eqref{initcond} fully fix $w_R$ to the right of the right shock, $x>v_rt$, and $w_L$ to the left of the left shock, $x < -v_l t$. Further, since the velocities are greater than 1, equations \eqref{bwc} and \eqref{wc} fully fix $w_C$ in the center region, $-t<x<t$. The remaining parts are the areas between the regions' interfaces and the shocks, $t<x<v_r t$ and $-v_l t<x<-t$. These are fixed by equations \eqref{restRL}, but only up to $O(g)$ terms. Hence, the order $g$ problem does have a unique solution. The space of solutions include the two-shock solution considered in paragraph \ref{sssectshocks}, which is that obtained by setting $w_R^+(x) = w_l +gw_l^2 + O(g^2)\ (x<0)$ and $\b w_L^-(x) = w_r + g w_r^2 + O(g^2)\ (x>0)$. But it also includes a continuum of solutions, which in general present four shocks, at $x=v_rt$, $x=-v_lt$ and also at $x=\pm t$.

Despite the non-uniqueness of the solution to order $g$, it is clear that the steady state is unique. Indeed, the steady state is completely determined by the center region, fixed by equations \eqref{bwc} and \eqref{wc} and the decomposition \eqref{wbw}. We find
\beq
	w_C = w_l -g w_l(w_r-w_l),\quad \b w_C = w_r + gw_r(w_r-w_l),
\eeq
and this leads to the following steady-state energy current and density:
\begin{eqnarray}\label{eq:steady}
j_s &=& w_C - \b w_C = (w_l-w_r)(1+ g(w_l+w_r)+O(g^2)),\\
h_s &=& w_C+\b w_C + g w_C\b w_C = w_l+w_r + g(w_l^2-w_lw_r+w_r^2) +O(g^2),
\end{eqnarray}
in agreement with the two-shock solution \eqref{twoshockssol}.

\subsubsection{Approach to the steady state and spreading of the shocks}

We now consider how the steady states is approached, and study the internal structure of the shocks. In order to do so, we extend beyond the first order in $g$ the analysis by using properties we expect to be valid at higher orders. The following results are conjectural.

{\em Approach to the steady state.} The steady state emerges at large time in the center region. In order to understand how the system approaches steadiness we have to look at the large time behavior of $w$ and $\bar w$ in this center region, say at $x\approx 0$. Let us solve \eqref{eq:linear} to all order in $g$. Continuing the solution in the central region all the way beyond the region, at $t=0$ we find an initial condition that is a sharp profile (modified by the jumps through the shocks). Because of this sharp profile, it is better to consider the space derivatives $w'$ and $\bar w'$, whose initial conditions are then of the form $w'(x,t=0)=\bar w'(x,t=0)\propto\delta(x)$. These have smooth Fourier transforms, and the solution can be written as
\[ w'(x\approx 0, t) \simeq \int dp \big[ U_p e^{i(px-E^+_pt)} + V_p e^{i(px-E^-_pt)}\big] ,\]
with $E^\pm_p$ the two branches of the dispersion relation in the center region and $U_p$ and $V_p$ smooth functions of $p$. In the bulk of the steady state, we assume that we can solve recursively, order by order in $g$, the system of differential equations by bringing it to linear equations. Thus the above form should also be valid, but with modified dispersion relation. This is our working hypothesis. Assuming that $U_p$ and $V_p$ stay smooth (because we are dealing with the derivative of $w$), for $x$ fixed the large time behavior of the integral representing $w(x,t)$ can be evaluated by a saddle point approximation. The saddle is at some $p_*$ and the saddle point integral is then of the form $\int dp\, e^{i\,{\rm const}\,(p-p_*)^2t}$. As a consequence, $w'(x,t)$ decreases as $1/\sqrt{t}$. This behavior has been seen in the approach to the steady state of the non-equilibrium Ising model \cite{Luca_Viti} and in Ising quantum quenches \cite{CEF}.

{\em Estimating the width of the shocks.} We consider a similar type of calculation in order to estimate the width of the shocks. For this purpose, let us assume that $E_p^\pm$ is of the form $E^\pm_p= \pm p(\alpha^\pm + \gamma^\pm p^2+\cdots)$, where the coefficients $\gamma^\pm$ are of order $g^2$. Although parity constrains the possible dispersion relations, we do not know how to justify within this picture the absence of a $p^2$ term in full generality, except that it is valid in the Ising case, but we will see that the result agrees with an independent argument based on quantum flows. 

We want to probe points at the ``shoulders'' of the shocks, so that we set $x\pm v_C^\pm t=\epsilon\, t$ and we look at the large time behavior with $\epsilon$ fixed but large enough. Again, we perform a saddle point approximation. A simple calculation gives the saddle point to be $p_\star\sim \sqrt{\ep}$, and the saddle phase is $e^{i\,{\rm const}\,\ep^{3/2}t}$. Replacing $\ep t$ by $x\pm v_C^\pm t$ leads to the phase $\exp\lt[i\,{\rm const}\lt(\frc{(x\pm v_C^\pm t)^3}t\rt)^{1/2}\rt]$, and thus we estimate the width of the shocks to be $\delta\xi_{\rm shock}\simeq (gt)^{1/3}$. This scaling relation is reminiscent to that appearing in KP or KdV equations \cite{Aba}.

Of course this result is conjectural because the argument is based on hypothesis that are not fully supported by calculations: the use of the linear approximation near the shock, and the form of the dispersion relation. Nevertheless we find the same estimate using a quite different approach in the following section.

\section{Quantum Virasoro flows and quantum hydrodynamics}

In this short Section we would like to make a step towards connecting the previous hydrodynamic formalism to possible quantum flows on Virasoro modules\footnote{This tentative formulation was motivated by a seminar given by J. Cardy in December 2014, at the conference ``Mathematical physics of non-equilibrium quantum systems", King's College London, in which he presented a possible connection between $T\bar T$ perturbations and CFTs in random metrics \cite{Cardy-London}. The approach we follow is different from that of J. Cardy, as we use a purely algebraic setup without appealing to Hubbard-Stratonovich transformations.}. 
The first step consists in reformulating the $T\bar T$ perturbations as random diffeomorphisms. This connection, based on the fact that the stress tensor is the generator of diffeomorphisms, is made here only to first order in $g$. We will use it to recover some of the previous results from the quantum flow perspective.

To be more specific, let $\phi(x)$ be a chiral primary field on the line -- with scaling dimension $h$ -- and consider its time evolution $\phi(x,t)$ with respect to the perturbed hamiltonian (\ref{eq:hamilton}). Because $\bar T$ commutes with $\phi$ and since $T$ is the generator of diffeomorphisms, we claim that the hamiltonian flow on chiral fields such as $\phi$  is that induced by the ``vector field'' $v(x) :=1+ g \bar T(x)$, that is:
\begin{eqnarray} \label{eq:H-flow}
 \phi(x,t) :=e^{-itH}\,\phi(x)\,e^{+itH} = [X_t'(x)]^h\ \phi(X_t(x)), 
 \end{eqnarray}
with $X_t$ the flow line generated by $v$ and started at $x$ at time $t=0$, i.e.
\begin{eqnarray} \label{eq:Vir-flow}
 \dot X_t(x)=v(X_t(x))=1+ g \bar T(X_t(x)) ,\quad \mathrm{with}\quad X_{t=0}(x)=x.
  \end{eqnarray}
The statement (\ref{eq:H-flow}) is actually valid to first order in $g$ only, because claiming that $\phi(x,t)$ commutes with $\bar T$ is true only to first order in $g$. 

To prove eq.(\ref{eq:H-flow}) we have to prove that $\phi(x,t)=[X_t'(x)]^h\ \phi(X_t(x))$ is solution of $\partial_t\phi(x,t)=-i[H,\phi(x,t)]$. Let us first recall well-known relations between diffeomorphisms, flow lines and the stress tensor. Since $T$ is the generator of diffeomorphisms, its commutation relation with a primary field $\phi$ of conformal dimension $h$ read 
\[ -i\big[ \int dy\, \xi(y)T(y)\,  ,\,\phi(x)\, \big] = h\, \xi'(x)\,\phi(x)+ \xi(x)\, \phi'(x).\]
Hence, for $\phi(x,t)=[X_t']^h\, \phi(X_t)$ we have
\[ -i\Big[\int dy\, \xi(y)T(y)\,,\,[X_t']^h\, \phi(X_t)\Big]= h\,\xi'(X_t)\, [X_t']^h\phi(X_t) + \xi(X_t)\, [X_t']^h\phi'(X_t),\]
provided that $X_t$ commutes with $T$. The latter commutator coincides with the time derivative $\partial_t\phi(x,t)$ provided we choose $X_t$ such that $\dot X_t=\xi(X_t)$, because then $\dot X_t'=X_t'\,\xi'(X_t)$ by the chain rule. That is: to solve for the time evolution we have to choose $X_t$ as the flow line started at $x$ associated to the vector field $\xi$. Now, our perturbed CFT hamiltonian is $H=\int dy \big( T(y)+\bar T(y) + g T(y)\bar T(y)\big)$. Since $\bar T$ commutes with any chiral field, we can apply the above remark to first order in $g$ provided we identify the vector field $\xi$ with $1+g\bar T$.

Thus, at least to first order in $g$, the dynamics of chiral fields is encoded into properties of the flows (\ref{eq:Vir-flow}) that we would like to call quantum Virasoro flows. These flows are random because $\bar T$ is a fluctuating data. We do not yet know whether these quantum flows can be made mathematically rigorous -- of course defining the probability distribution functions of the flow trajectories from data on the Virasoro algebra requires appropriate regularization. 

Let us nevertheless use this flow to deduce properties that were found in the previous sections.

The temperature dependent deformation of the light cone is easy to understand using quantum Virasoro flows. The light cone velocity 
is simply the mean velocity  $\bra 1+g\bar T(X_t)\ket$, evaluated in the initial state. It is equal to $v_{l,r}=1 + g\frac{c\pi}{12} T^2_{l,r}$, with $T_{l,r}$ the left/right temperatures, depending in which region the shock front is moving.

The quantum Virasoro flows also give an alternative understanding (based on scaling arguments) for the $t^{1/3}$-scaling behavior we have been arguing for in the previous Section. Indeed let us consider two (quantum) trajectories started at neighborhood points $x$ and $x+\delta x$, and let $\delta X_t$ be their separation after a time duration $t$. By construction, its time evolution is $\dot{(\delta X_t)}= g \, \delta \bar T(X_t)$. Since $\bar T$ is of scaling dimension $2$, $\delta \bar T$ typically scales like $1/(\delta X_t)^2$, or equivalently $\dot{(\delta X_t)}\sim g/(\delta X_t)^2$. By integration this yields that $(\delta X_t)^3$ scales linearly in time,  $(\delta X_{t})^3\sim g\, t$, which indicates that the correlation length in quantum Virasoro flows scales with time as $(g\, t)^{1/3}$, which thus gives the natural width of the shocks. It would be interesting to make such arguments less conjectural.

\section{Conclusion}

In this paper, we reviewed the general hydrodynamic approach to non-equilibrium quantum field theory. In order to study non-equilibrium effects it was important to keep two conserved quantities, including a parity-odd conserved charge, in the local fluid description. We then applied these ideas to develop a hydrodynamic approach for conformal field theory perturbed by the $T\b T$ operator. We verified by direct quantum computations that the main aspects of hydrodynamics emerge at first nontrivial order in the coupling $g$.

An immediate question is about the higher orders in perturbation theory. Clearly there were many simplifications at first order, in particular the emergence of Lorentz invariance, which are not present at higher orders. At higher orders, the fact that the perturbation is irrelevant may become more crucial and generate non-universal contributions.

It will also be interesting to study the internal structure of the shocks, already at first order in $g$. We believe that the conjectured scaling relations are correct and universal. An important further question concerns the existence or not of universal structures inside the shocks, in a way similar to what has been revealed in classical fluctuating hydrodynamics \cite{Spohn}.

In this paper we have focussed on  mean transport phenomena and it will be very interesting to describe fluctuations and their large deviation functions. In particular one may wonder whether the extended fluctuation relations \cite{BD13xx}, proved in pure non-equilibrium conformal field theory and in integrable systems (and conjectured to hold as well in higher-dimensional CFT \cite{Nat-Phys}), remain valid once the $T\bar T$ perturbation has been turned on.

\bigskip

{\bf Acknowledgements:} 
We are pleased to thank J. Cardy for motivating discussions as well as A. Abanov, E. Akkermans, E. Boulat and A. Lamacraft for sharing with us their insights. D.B. thanks the Kadanov Center for Theoretical Physics at the University of Chicago for hospitality where part of this work was done. B.D. thanks J. Bhaseen, A. Lucas and K. Schalm for discussions and collaborations on this subject. D.B. and B.D. thank the Institute of Physics of the University of Amsterdam for hospitality during the LDQCM 2015 Workshop, 29 June - 3 July 2015. This work was supported in part by the ANR contracts ANR-2010-BLANC-0414 and ANR-14-CE25-0003-01.

\appendix

\section{Relativistic thermodynamics} \label{apprel}

Consider a one-dimensional quantum system with the following general structure: a hamiltonian $H$ generating time translations and a momentum operator $P$ generating space translations, both expressed as integrals of local densities:
\beq\label{densh}
	H = \int \dd x\,h(x),\quad P = \int \dd x\,p(x);
\eeq
and the conditions that (1) both are conserved:
\beq
	\p_t h + \p_x j =0,\quad \p_t p + \p_x k=0,
\eeq
and that (2) the momentum density is equal to the energy current, plus possibly an even derivative of the energy current:
\beq
	p = j + \mu\, \p_x^{2\ell} j,\quad \ell\geq1
\eeq
where $\mu$ is some constant. We are interested in the ``generalized thermal'' state, where both the hamiltonian and the momentum operator are present. We parametrize it by a ``rest-frame temperature'' $T$ and a ``boost parameter'' $\theta$:
\[
	\bra{\cdots}\ket_{\theta} = \frc{\Tr\lt( e^{-\frc1T(\cosh\theta \,H - \sinh\theta \,P)} \cdots\rt)}{\Tr \lt(e^{-\frac1T (\cosh\theta \,H - \sinh\theta \,P)}\rt)},
\]
and below we will denote $\hh_\theta = \bra h\ket_\theta$, etc.

We assume that there is parity symmetry whereby $H$ is invariant and $P$ changes sign, that the ground state is parity invariant, and that the limit $T\to0$ of the state $\bra\cdots\ket_\theta$ is the ground state, independently of $\theta$. Further, we assume that the densities are homogeneous and (as the notation suggests) not explicitly time dependent: for instance, $[P,h(x)] = i\p_x h(x)$ and $[H,h(x)] = -i\p_t h(x)$. Note that there is a simple gauge symmetry: we may change $h(x)\mapsto h(x)+\alpha{\bf 1}$ and $k(x)\mapsto k(x)+\alpha'{\bf 1}$ for any constants $\alpha$ and $\alpha'$ without changing any of the conditions. We will also use the principle according to which
\beq\label{princ}
	[P,\Or]=0\; \Rightarrow\; \Or\propto{\bf 1}
\eeq
for local densities $\Or$.  Note that $\pp_\theta = \jj_\theta$ since the average of a derivative of a homogeneous density is zero.

It turns out that this structure is constraining enough to imply relativistic invariance. We will indeed show from the above that averages in a generalized thermal state take their relativistic form, and that a standard thermodynamic relation between the variations of pressure and of temperature hold: under the choice of gauge such that $\hh_0+\uu_0 \to 0$ as the temperature $T$ goes to 0,
\beqa
	{\hh}_\theta &=& \cosh^2\theta \,\hh_0 + \sinh^2\theta\,\uu_0\n
	{\uu}_\theta &=& \sinh^2\theta \,\hh_0 + \cosh^2\theta\,\uu_0\n
	{\pp}_\theta &=& \sinh\theta\cosh\theta\,\big(\hh_0 + \uu_0\big)
	\label{p1}
\eeqa
and
\beq\label{p2}
	(\hh_0 + \uu_0)\dd T = T\dd \uu_0.
\eeq
Remark that, from \eqref{p2}, if $\hh_0 \sim A\,T^q$ for some $q>1$ as $T\to0$, then $\uu_0 \sim A'\,T + AT^q/(q-1)$, where $A$ and $A'$ are some constants\footnote{In fact, there is a stronger statement, based on a weaker assumption: if $\hh_0 \leq A\,T^q\ \forall\ T>0$ small enough, then $\uu_0 = A'\,T + g(T)$ with $g(T)\leq \frc{Aq}{q-1} T^q\  \forall\ T>0$ small enough.} (the case $q\leq 1$ would pose physical problems as it implies that the pressure becomes negative in some temperature range). Equations \eqref{p1} can also be written in the Lorentz-invariant form $\TT^{\mu\nu}_\theta = \uu_0\,\eta^{\mu\nu} + (\hh_0+\uu_0)\,u^\mu_\theta u^\nu_\theta$, where $\eta^{\mu\nu} = {\rm diag}(-1,1)$ is the Minkowsky metric, $\TT^{\mu\nu}_\theta = \mato \hh_\theta & \pp_\theta \\ \pp_\theta & \uu_\theta \matf$ is the stress-energy tensor and $u^\mu_\theta = \mato \cosh\theta \\ \sinh\theta\matf$ is the relativistic velocity.

These are purely dynamical relations. All the information about the particular model at hand may be imbedded into the ``equation of state'', and once this information is given, the above provide all averages explicitly in terms of the temperature and the boost parameter.
By parity invariance in the thermal state, ${\pp}_0=0$. Hence there is one parameter characterizing ${\hh}_0$ and ${\uu}_0$, so we can define the function $F$ such that the following equation of state holds:
\beq
	{\uu}_0 = F({\hh}_0).
\eeq
Then, given $F$ the relation \eqref{p2} provides the explicit temperature dependence in the thermal state,
\beq\label{Tdep}
	\log T = \int^{\uu_0} \frc{\dd \ell}{\ell + F^{-1}(\ell)}
	= \int^{\hh_0} \frc{\dd \ell\,F'(\ell)}{\ell+F(\ell)}.
\eeq
Note that in the generalized thermal state, the equation of state is more complicated, and, from relations \eqref{p1}, takes the form
\beq
	\sqrt{\big(\hh_\theta+\uu_\theta\big)^2 - 4\pp_\theta^2}
	-\hh_\theta + \uu_\theta = 2F\lt(
	\sqrt{\big(\hh_\theta+\uu_\theta\big)^2 - 4\pp_\theta^2}
	+\hh_\theta - \uu_\theta\rt).
\eeq

Although we restricted the analysis to one-dimensional systems, the above hold as well in higher dimensions, where the effective one-dimensional system is obtained by averaging on hyper-surfaces transverse to the $x$ direction in a $d$-dimensional space. Denote by $\TT^{\mu\nu}$ the current associated with invariance under translation in the direction $x^\nu$, with $x^0=t$, $x^1=x$ and transverse direction $x^\perp = (x^2,\ldots,x^d)$. Then $\hh(x)$, $\pp(x)$ and $\uu(x)$ are transverse averages of the energy density $V^{-1}\int_V \dd^\perp x\,\TT^{00}(x)$, of the $x$-momentum density $V^{-1}\int_V \dd^\perp x\,\TT^{01}(x)$, and of the $x$-pressure density $V^{-1}\int_V \dd^\perp x \,\TT^{11}(x)$, respectively, where $V$ is the transverse volume. This is particularly relevant for energy transport, as the direction of transport makes the system effectively one-dimensional. We note that, for instance, in scale invariant systems in $d$ dimensions, the equation of state is
\beq
	{\uu}_0 = d \,{\hh}_0
\eeq

{\bf Proofs}

First we may redefine the energy density and current as follows:
\beq
	\t h= h +\mu\, \p_x^{2\ell} h,\quad \t j = j + \mu \,\p_x^{2\ell} j.
\eeq
This is such that $\t h$ is still an energy density, $H = \int \dd x\, \t h$ up to local densities at infinity, that the conservation equation holds, $\p_t \t h + \p_x \t j=0$, and that the new energy current is exactly equal to the momentum density, $\t j=p$. Further, these re-definitions do not affect the properties of the energy density and current under a parity transformation. Hence without loss of generality, below we assume that $h$ and $j$ have been chosen in such a way that $j=p$.

For lightness of notation we omit the index $\theta$, except when it is equal to 0 (thermal state). First we show that there exists a boost operator $B$ giving rise to the algebra
\beq\label{alg}
	[B,H] = P,\quad [B,P]=H,\quad[H,P]=0.
\eeq
Indeed, consider
\beq
	B = -i\int \dd x\,xh(x).
\eeq
Then we have
\[
	[B,H] = -i\int \dd x\,x[h(x),H] = \int \dd x\,x\p_t h(x)
	= -\int \dd x\,x\p_xp(x) = \int \dd x\,p(x) = P
\]
and
\[
	[B,P] = -i\int \dd x\,x[h(x),P] = -\int \dd x\,x\p_x h(x)
	= \int \dd x\,h(x) = H,
\]
up to local densities at infinity. Local densities at infinity do not contribute whenever the operators are exponentiated either when acting on local observables by adjoint action, or in some density matrix where averages of local observables are taken\footnote{That is, this is valid from the point of view of the properties of time and space translations or of states, not the point of view of the averages of $H$ and $P$ in a state.}. \hfill $\square$

Remark that we used the fact that the energy current is the momentum density only in the first calculation: for any translation invariant system we have $[B,P]=H$, even without relativistic invariance.  

Naturally, the algebra \eqref{alg} implies that the generalized thermal state is a boosted state, so that
\[
	{\Or}_\theta = \bra e^{\theta B}\,\Or\, e^{-\theta B}\ket_0
\]
for any local observable $\Or$.

We next show that
\beq\label{tp1}
	\p_\theta \hh  = 2\pp,\quad
	\p_\theta\pp = \hh+\uu,\quad \p_\theta \uu = 2\pp
\eeq
under the choice of gauge characterized by the fact that $\hh_0+\uu_0\to0$ as $T\to0$. 

Let $B(x)=e^{-iPx}\, B\, e^{iPx}$ be the translated of $B$. We have $B(x)=B+ixH$.  Let now $[B,h](x) = e^{-iPx}[B,h]e^{iPx}$ be the translated of $[B,h]$ with $h=h(0)$. Since $[B,h](x)=[B(x),h(x)]$ we have:
\[ [B,h](x)= [B,h(x)]-x\partial_xp(x),\]
where we use the equation of motion $i[H,h(x)]=-\partial_x p(x)$.
%
%
Integrating,
\[
	P = [B,H] 
	= \int\dd x\,(-p(x) + [B,h](x))
	= -P + \int\dd x\,[B,h](x)
\]
which is valid up to local densities at infinity, whereby
\[
	[B,h](x) = 2p(x) + i \p_x b(x)
\]
for some local density $b(x)$.

Let us show that $b(x)$ is a homogeneous density, $\p_x b(x) = -i[P,b(x)]$. By definition $[B,h](x)$ is a homogeneous density: $-i[P,[B,h](x)] = \p_x[B,h](x)$. Since also $p(x)$ is, then $\p_x b(x)$ is a local homogeneous density:  $\p_x [P,b(x)] = i\p_x^2 b(x)$. The last equation means that $[P,b(x)] - i\p_x b(x)$ is independent of $x$. According to \eqref{princ}, the only local densities that are independent of $x$ are those proportional to the identity operator ${\bf 1}$, hence
\beq\label{pf3}
	[P,b(x)]-i\p_xb(x) = \alpha {\bf 1}
\eeq
for some constant $\alpha$. Further, under parity we have that $[B,h](x) - 2p(x)$ changes sign (in addition to $x$ changing sign). Hence $b(x)$ changes sign, again up to a local density that is independent of $x$. That is, under parity, $b(x)\mapsto -b(-x) + \t\alpha{\bf 1}$ for some other constant $\t \alpha$. Putting this into \eqref{pf3}, we find that the left-hand side changes sign under parity, whereby we must have $\alpha=0$. Therefore
\beq\label{Bh}
	[B,h](x) = 2p(x) +[P,b(x)].
\eeq
and this leads to the first of \eqref{tp1}.

By a similar argument,
\[
	[B,p(x)] = x\p_x k(x) + [B,p](x)
\]
and integrating,
\[
	H = [B,P] = \int \dd x\,(-k(x) + [B,p](x))
\]
up to local densities at infinity, therefore
\beq\label{Bp}
	[B,p](x) = h(x) + k(x)+i \p_x b'(x)
\eeq
for some local density $b'(x)$. Again $\p_x b'(x)$ is a homogeneous density, so that, by arguments as above, $[P,b'(x)]-i\p_xb'(x) = \alpha'{\bf 1}$ for some constant $\alpha'$. We cannot use parity symmetry in order to fix the constant $\alpha'$, however we may absorb it into an appropriate gauge choice for the sum of the energy density $h(x)$ and the pressure $k(x)$. With this choice of gauge, the second of \eqref{tp1} follows. Since as $T\to0$ the state specializes to the ground state, which is assumed to be Lorentz invariant, then from \eqref{tp1} this choice of gauge can be characterized by the fact that $h+k\to0$ as $T\to0$.

Finally, let us set $x=0$ in \eqref{Bp}, giving
\[
	[B,p]= h+k+[P,b']-\alpha'{\bf 1}.
\]
We apply $[H,\cdot]$ on both sides and use the algebra \eqref{alg} as well as the conservation equations \eqref{eq:conserve}. We obtain on the left-hand side
\[
	-[P,p] + [B,[H,p]] = -[P,p] + [B,[P,k]] = [P,[B,k]-p] + [H,k]
\]
and on the right-hand side
\[
	[H,h]+[H,k]+[P,[H,b']] = [P,p]+[H,k]+[P,[H,b']].
\]
Comparing, and using \eqref{princ}, we get the relation
\[
	[B,k] = 2p + [H,b'] + \alpha'' {\bf 1}
\]
for some constant $\alpha''$. This gives $\p_\theta \uu = 2\pp + \alpha''$. Again, as $T\to0$ the state specializes to the ground state which is Lorentz invariance, whence $2\pp_0 + \alpha''=0$. Further, by parity invariance of the ground state $\pp_0=0$, whereby $\alpha''=0$. This shows the third relation of \eqref{tp1}. \hfill $\square$

From \eqref{tp1}, relations \eqref{p1} immediately follow: keeping the temperature $T$ fixed, the system of three first-order ordinary differential equations \eqref{tp1} in $\theta$ has a unique solution with the initial conditions $\hh_0$, $\pp_0$ and $\uu_0$ at $\theta=0$, and one can check that $\eqref{p1}$ is a solution. \hfill $\square$

Finally we show \eqref{p2}. For this purpose, we show the two relations
\beqa
	\sinh\theta\,\p_\theta \pp - \cosh\theta\, \p_\theta\hh &=&
	\sinh\theta\,T\p_T\hh - \cosh\theta\,T\p_T\pp \n
	\sinh\theta\,\p_\theta \uu - \cosh\theta\, \p_\theta\pp &=&
	\sinh\theta\,T\p_T\pp - \cosh\theta\,T\p_T\uu\label{tp2}
\eeqa
Denote the density matrix as $e^{-\beta_h H -\beta_p P}$. These two relations are a re-writing of
\[
	\frc{\p}{\p\beta_h} \pp= \frc{\p}{\p\beta_p} \hh,\quad
	\frc{\p}{\p\beta_h} \uu= \frc{\p}{\p\beta_p} \pp
\]
respectively. The first one is trivial as the equality boils down to $\int \dd x\,\genc{h(x)p(0)} = \int \dd x\,\genc{h(0)p(x)}$ where $\genc{\cdots}$ is the connected average. The second one in \eqref{tp2} can be obtained by differentiating the first one with respect to $\theta$ (assuming that the $T$ and $\theta$ derivatives can be interchanged), and by using \eqref{tp1}. Then consider the second relation of \eqref{tp2} at $\theta=0$, which is $\p_\theta\gen\pp|_{\theta=0} = T\p_T\gen\uu_0$. Using \eqref{tp1} this is $\gen\hh_0 + \gen\uu_0 = T\p_T \gen\uu_0$. \hfill $\square$

\end{document}